\documentstyle[12pt]{article}
\begin{document}
\centerline{\large{\bf On the importance of the Bohmian approach for}}
\vskip 0.2in
\centerline{\large{\bf interpreting CP-violation experiments}}

\vskip 0.5in

\centerline{Dipankar Home\footnote{e-mail:dhom@boseinst.ernet.in}}
\vskip 0.2in
\centerline{Bose Institute, Calcutta 700009, India}
\vskip 0.2in
\centerline{A.S.Majumdar\footnote{e-mail:archan@boson.bose.res.in}}
\vskip 0.2in
\centerline{S.N.Bose National Centre for Basic Sciences}
\vskip 0.1in
\centerline{Block-JD, Sector III, Salt Lake, Calcutta 700091,
India.}

\vskip 2.0in

\begin{abstract}

We argue that the inference of CP violation in experiments
involving the $K^0 - \bar{K^0}$ system in weak interactions of
particle physics is facilitated by the assumption of particle trajectories
for the decaying particles and the decay products. A consistent
explanation in terms of such trajectories is naturally incorporated
within the Bohmian interpretation of quantum mechanics.

\end{abstract}

\pagebreak

\noindent
I. Introduction

The Bohm model is able to provide a causal interpretation of
quantum mechanics in a consistent manner~[1]. At the same time, the
predictions of Bohmian mechanics are in exact agreement with the
standard quantum mechanical predictions for observable
probabilities in all usual experimental situations.
In this paper we shall be concerned
with examining the possible importance of the Bohmian approach in
interpreting certain experiments whose understanding in terms of
the standard interpretation is rather ambiguous.

For the purpose of {\it reinterpreting} the standard quantum formalism
using the Bohmian scheme, a wave function $\psi$ is {\it not} taken
to provide a complete specification of the state of an individual
system; an {\it additional} ontological ``position'' coordinate (an
objectively real ``position'' existing irrespective of any external
observation) is
ascribed to an individual particle. The ``position'' coordinate of
the particle evolves with time obeying an equation which can be
derived from the Schrodinger equation (considering
the one dimensional case)
\begin{eqnarray}
i\hbar {\partial\psi \over \partial t} = H\psi \equiv - {\hbar^2
\over 2m} {\partial^2 \psi \over \partial x^2} + V(x)\psi
\end{eqnarray}
by writing
\begin{eqnarray}
\psi = Re^{iS/\hbar}
\end{eqnarray}
and using the continuity equation
\begin{eqnarray}
{\partial \over \partial x} (\rho v) + {\partial\rho \over \partial
t} = 0
\end{eqnarray}
for the probability distribution $\rho(x,t)$ given by
\begin{eqnarray}
\rho = \vert \psi \vert^2.
\end{eqnarray}
It is important to note that $\rho$ is ascribed an {\it
ontological}
significance by regarding it as representing the probability
density of ``particles'' occupying {\it actual} positions. In
contrast, in the standard formulation $\rho$ is interpreted as the
probability density of {\it finding} a particle around a certain
position. Setting ($\rho v$) equal to the quantum probability current
leads naturally to the Bohmian interpretation whrere the particle
velocity $v(x,t)$ is given by
\begin{eqnarray}
v \equiv {dx \over dt} = {1\over m}{\partial S \over \partial x}
\end{eqnarray}
The particle ``trajectory''
is completely deterministic  and is obtained by integrating (5)
with the appropriate initial conditions.
The essential significance
of Bohm's model lies in providing an elegant solution to the
measurement problem (which has been described by Weinberg~[2] as
``the most important puzzle in the interpretation of quantum
mechanics'') without requiring wave function collapse, since
according to the Bohmian interpretation, in any measurement a
definite outcome is singled out by the relevant ontological
position coordinate.

In view of the importance of the Bohm model in providing not only
an internally consistent {\it alternative} interpretation of the
standard quantum formalism, but also perhaps the neatest solution
to the measurement problem~[1], it should be worthwhile to look for
specific situations where the {\it conceptual superiority} of Bohm's
model over the standard interpretation may become easily
transparent. To this end, we now proceed to examine the analysis of
a fundamentally important experiment of particle physics, namely,
the discovery of CP-violation~[3].

\noindent
II. The CP-violation experiment

C(charge conjugation) and P(parity) are two of the fundamental
discrete symmetries of nature, the violations of which have not
been empirically detected in phenomena other than weak interactions.
If a third discrete symmetry T(time reversal) is
taken into account, there exists a fundamental theorem of quantum
field theory, viz., the CPT theorem which states that all physical
processes are invariant under the combined operation of CPT.
Nevertheless, there is no theorem forbidding the violation of CP
symmetry, and indeed, there have been several experiments to
date~[4], starting from the pioneering observation of Christenson,
Cronin, Fitch and Turlay~[3], that have revealed the occurrence of CP
violation through weak interactions of particle physics involving
the particles $K^0$ and $\bar{K^0}$. The eigenstates of strangeness
$K^0$ \hskip 0.1in $(s=+1)$ and its CP conjugate $\bar{K^0}$ \hskip
0.1in $(s=-1)$ are produced in strong interactions, for example,
the decay of $\Phi$ particles. Weak interactions do not conserve
strangeness, whereby $K^0$ and $\bar{K^0}$ can mix through
intermediate states like $2\pi, 3\pi, \pi\mu\nu, \pi e\nu$, etc. The
observable particles, which are the long lived $K$-meson $K_L$, and
the short lived one $K_S$, are linear superpositions of $K^0$ and
$\bar{K^0}$, i.e.,
\begin{eqnarray}
\vert K_L \rangle & = & (p\vert K^0\rangle - q\vert
\bar{K^0}\rangle )/ \sqrt{\vert p\vert^2 + \vert q\vert^2} \\
\vert K_S \rangle & = & (p\vert K^0\rangle + q\vert
\bar{K^0}\rangle )/ \sqrt{\vert p\vert^2 + \vert q\vert^2}
\end{eqnarray}
which obey the exponential decay law $\vert K_L\rangle \rightarrow
\vert K_L\rangle exp(-\Gamma_L t/2)exp(-im_Lt)$ and analogously for
$\vert K_S\rangle$, where $\Gamma_L$ and $m_L$ are the decay width
and mass respectively of the $K_L$ particle. It follows from (6)
and (7) that
\begin{eqnarray}
\langle K_L\vert K_S\rangle =  {|p|^2-|q|^2 \over |p|^2+|q|^2}
\end{eqnarray}

CP violation takes place if the states $\vert K_L\rangle$ and
$\vert K_S\rangle$ are not orthogonal. Through weak interactions
the $K_S$ particle decays rapidly into channels such as $K_S
\rightarrow \pi^{+}\pi^{-}$ and $K_S \rightarrow 2\pi^{0}$  with a
mean lifetime of $10^{-10}s$, whereas, the predominant decay modes
of $K_L$ are $K_L \rightarrow \pi^{\pm}e^{\pm}\nu$ (with branching
ratio $\sim 39\%$), $K_L \rightarrow \pi^{\pm}\mu^{\pm}\nu (\sim
27\%)$, and $K_L \rightarrow 3\pi ( \sim 33\%)$ [4]. The CP
violating decay mode $K_L \rightarrow 2\pi$ is extremely rare (with
branching ratio $\sim 10^{-3}$) in the background of the other
large decay modes. Considering the Schrodinger evolution, if the
analysis of the term corresponding to $K_s$ in the relevant initial wave
function shows that it cannot contribute significantly to the emission
of two pions with suitable momenta and locations, then one can infer the
occurrence of CP violation in this particular situation. In other words
such $2\pi$ can only arise through the $K_L$ decay mode. The momenta and
locations of the emitted pions are important since
the key experimental issue is to detect the $2\pi$
particles coming from the decay of $K_L$ and identify them
as coming from $K_L$ and {\it not} $K_S$.

In a typical experiment to detect CP violation, an initial state of
the type
\begin{eqnarray}
\vert\psi_i\rangle = (a\vert K_L\rangle + b\vert K_S\rangle )
\end{eqnarray}
is used which is a coherent superposition of the $K_L$ and $K_S$
states. Such a state has been produced by the technique of
`regeneration'~[5] which has been used in a large number of
experiments~[6]. The common feature of all these experiments is the
measurement of the vector momenta $\stackrel{\rightarrow}{p_i}$ of
the charged decay products $\pi^{+}\pi^{-}$ or $2\pi^0$ from the
decaying pions. It is only the {\it type} of instrument used for actually
measuring the momenta that varies from experiment to experiment.

\noindent
III. Bohmian trajectories

To see how the Bohmian interpretation helps in drawing the
relevant inference from this experiment, we concentrate on the
analysis of a single event in which the two emitted pions from a
decaying kaon are detected by two detectors respectively along two
different directions. From the measured momenta
$\stackrel{\rightarrow}{p_1}$ and $\stackrel{\rightarrow}{p_2}$,
the ``trajectories'' followed by the individual pions are
{\it retrodictively} inferred {\it assuming} that they have followed {\it
linear ``trajectories''}.  The point of intersection of these
retrodicted ``trajectories'' is inferred to be the point from which
the decay products have emanated from the decaying system; in other
words, what is technically known as the ``decay vertex'' is
determined in this way. The value of the momentum of the decaying
kaon is obtained by $\stackrel{\rightarrow}{p_k} =
\stackrel{\rightarrow}{p_1} + \stackrel{\rightarrow}{p_2}$. Once
the decay vertex and the kaon momentum is known, one estimates the
time taken by the kaon to reach the decay vertex from the source,
{\it again} using at this stage the idea of a {\it linear
``trajectory''}. If this time turns out to be much larger than the
$K_S$ mean lifetime ($\sim 10^{-10}s$), one infers that the
detected $2\pi$ pair must have come from $K_L$, which, as already
mentioned, is the signature of CP violation.

It is thus evident from the above discussion that the assumption
of a {\it linear ``trajectory''} of a freely evolving particle
(kaon or pion) provides a consistent explanation in support of
CP violation in such
an experiment. Within the standard interpretationm of quantum
mechanics, there is no way one can justify assigning a
``trajectory'' to a freely evolving particle. Moreover, assuming
such a ``trajectory'' to be {\it linear} is an additional {\it ad
hoc} input. One possible argument could be to assign localized wave
packets to emitted pions and kaons, and to use the fact that their
peaks follow classical trajectories in the case of a free
evolution. However, in the standard quantum mechanical description
of decay processes, the decay products are regarded as asymptotically
free, and hence should be represented by
plane wave states. Moreover, even if they are
approximated in some sense by localized wave packets, there would
be inevitable spreading of the wave packets. Even if this spreading
is regarded as negligible within the time interval concerned, a
`literal identification' of the wave packet with the particle
is conceptually impermissible without an {\it additional}
input at the fundamental level in the form of the notion of a
``particle'' with a definite position even when unobserved
(``particle'' ontology).

On the other hand, the assumption of {\it linear ``trajectories''}
followed by the decaying particles and the decay products is
amenable to a natural explanation within the Bohmian framework.
The decaying kaons as well as the
asymptotically free decay products are represented by plane
waves
\begin{eqnarray}
\psi \sim e^{ikx}.
\end{eqnarray}
Hence it follows that in the Bohmian scheme the velocity equation
(5) is in this case given by
\begin{eqnarray}
{dx \over dt} = {\hbar k \over m}
\end{eqnarray}
which when integrated provides the linear ``trajectories'' of the
particles. These trajectories are ontological
and deterministic. Therefore, in this interpretation, the
exact position coordinates of the ``decay vertex'' can be assigned
in a natural way by retrodicting the pion ``trajectories'' without any
inconsistencies of the type inherent in the standard
interpretation. Hence, it seems necessary that the standard
formalism of quantum mechanics needs to be supplemented with the
Bohmian interpretation of ontological particle ``trajectory'' (in
the sense that the particle has traversed a well defined path even
when unobserved) to enable for the consistent inference of the
observation of CP violation in the actual experiments involving
kaon decays.

\noindent
IV. Concluding remarks

The main reasons for choosing, in particular, the CP violation
experiment for this purpose are the following. First, {\it unlike}
other common high energy experiments this particular experiment
involves {\it not} merely the measurement of some physical
quantities but {\it inferring} from the measured quantities the
violation of a fundamental symmetry property of the pertinent
physical interactions. Secondly, again unlike other common high energy
experiments, the effects of particle creation and annihilation are not
relevant for the important part of the experiment involved with the
prediction of CP violation, and no second quantized treatment is
required for the theoretical framework. The crucial phenomena of
particle decays which this experiment is concerned with, is
appropriately described in terms of the Schrodinger equation (see [4]
and references therein) for which there exists a consistent Bohmian
interpretation. Note that ignoring interpretational
nuances, if one tries to follow a very pragmatic approach and {\it
approximates} the plane wave states of the decay products by wave
packets whose peaks follow classical trajectories with finite
speeds, careful estimates need to be done to {\it quantify} the
resulting errors or fluctuations due to spreading of wave packets
by taking into account the {\it
actual} distances involved in the performed experiments. (Of course, the
estimates of these distances related to the particle trajectories are
fundamental from the Bohmian perspective.) This is
important because the CP violation effect is exceedingly {\it
small}; the branching ratio of the CP violating decay mode $K_L
\rightarrow 2\pi$ is $10^{-3}$. In {\it none} of the CP violation
experiments performed to date has this point been considered in the
relevant analysis.

We conclude by noting that this analysis suggests that it should be
worthwhile to look for more such appropriate examples where the
inadequacy or ambiguity of the standard formalism in
comprehending the results of the concerned experiments
can be avoided by using
the Bohmian interpretation. It should be appreciated that since there is
no measurement problem in the Bohmian interpretation, a Bohmian analysis
is useful for all experiments in quantum mechanics, and in particular
scattering experiments where it is required to know why particles are
detected where they are at the end of the experiment. The answer to this
is clear from the Bohmian perspective---the particles are detected where
they {\it actually are}. However, from the viewpoint of the standard
interpretation the explanation is rather obscure, as long as the
Schrodinger wave function is regarded as the complete description of the
physical system. In this context it has been recently argued~[7] that the
concept of quantum probability current, a full understanding of which is
provided by Bohmian mechanics, is fundamental for a genuine understanding
of scattering phenomena. Apart from this, it has been claimed~[8]
that a special significance of Bohmian mechanics lies in experiments
related to the measurement of time of flight of particles, and tunelling
time in particular for which it is difficult to find a consistent or
unambiguous definition within the standard framework of quantum
mechanics. All this is of course {\it different} from
empirically verifying a new consequence, if any exists, of the Bohmian
interpretation which is not obtainable from the standard
interpretation. Nevertheless, such investigations like the one
reported in this paper could be helpful in understanding more
clearly the relative merits of the standard and Bohmian
interpretations.

\vskip 0.5in

This work was supported by the Department of Science and
Technology, India.

\pagebreak

{\bf REFERENCES}

\vskip 0.5in

\begin{description}

\item[[1]]
P.R.Holland, ``The Quantum Theory of Motion'', (Cambridge
University Press, London, 1993);
D.Bohm, Phys. Rev. {\bf 85} (1952) 166; D.Bohm and
B.J.Hiley, ``The Undivided Universe'', (Routledge, London, 1993);
J.T.Cushing, ``Quantum Mechanics -
Historical Contingency and the Copenhagen Hegemony'', (University
of Chicago Press, Chicago, 1994).
\item[[2]] S.Weinberg, ``Dreams of a final theory'', (Vintage,
London, 1993) p. 64.
\item[[3]] J.H.Christenson, J.W.Cronin, V.L.Fitch and R.Turlay,
Phys. Rev. Lett. {\bf 13} (1964) 138.
\item[[4]] For a review, see for instance, K.Kleinknecht, in ``CP
violation'', edited by C.Jarlskog, (World Scientific, Singapore,
1989) pp. 41 -104.
\item[[5]] A.Pais and O.Piccioni, Phys. Rev. {\bf 100} (1955) 1487.
\item[[6]] For eample, see, C.Geweniger et al., Phys. Lett. B {\bf
48} (1974) 487; V.Chaloupka et al., Phys. Lett. B {\bf 50} (1974)
1; W.C.Carithers et al., Phys. Rev. Lett. {\bf 34} (1975) 1244;
N.Grossmann, et al., Phys. Rev. Lett., {\bf 59} (1987) 18.
\item[[7]] M.Daumer, D.Duerr, S.Goldstein and N.Zanghi, J. Stat.
Phys. {\bf 88} (1997) 967.
\item[[8]] C.R.Leavens, Phys Lett. A{\bf 197} (1995) 88; in ``Bohmian
Mechanics and Quantum Theory: An Appraisal'', eds. J.T.Cushing,
A.Fine and S.Goldstein (Kluwer, Dordrecht, 1996).

\end{description}

\end{document}